\documentclass[12pt,notitlepage,letterpaper]{article}
\usepackage{graphicx}
\usepackage{amsmath}
\usepackage{amssymb}
\usepackage{latexsym}
\usepackage{xcolor}
\usepackage{multirow}
\usepackage[round,sort&compress,comma,numbers]{natbib}
\usepackage[small,bf]{caption}
\usepackage[margin=1in]{geometry}

\makeatletter \renewcommand{\@biblabel}[1]{#1.} \makeatother

\graphicspath{%
{/Users/vitaly/research/pictures/}%
{figs/}%
}

\begin{document}

\title{Physical and mathematical modeling in experimental papers:
  achieving robustness of mathematical modeling studies}

\author{%
  {Vitaly V. Ganusov }\\
  {\small Department of Microbiology, University of Tennessee,
    Knoxville, TN 37996, USA}\\
  {\small Department of Mathematics, University of Tennessee,
    Knoxville, TN 37996, USA}\\
  {\small National Institute for Mathematical and Biological
    synthesis, Knoxville, TN 37996, USA}
}

\maketitle

\begin{abstract}

  Development of several alternative mathematical models for the
  biological system in question and discrimination between such models
  using experimental data is the best way to robust conclusions.
  Models which challenge existing theories are more valuable than
  models which support such theories.

\end{abstract}

Mathematical modeling has been especially strong in physical sciences
where theories, based on basic principles, were able to predict
energies associated with mass (famous equation $E=mc^2$) and existence
of fundamental particles (e.g., positron). There have been many
applications of mathematical modeling in biology but because it has
been harder to come up with basic principles, predictive successes of
mathematical theory-based predictions in biology are more modest than
that in physics. With the development of novel experimental methods to
collect unique data spanning different levels of organization, from
gene sequences and intracellular processes in individual cells to
populations of organisms, there has been an impressive rise in the
number of mathematical modeling studies in biology. Bioinformatics,
commonly defined as a science dealing with analysis of sequencing data
(but see \cite{Hogeweg.pcb11}), and systems biology, commonly defined
as a science dealing with intracellular gene networks, are two most
well recognized advances driven by the availability of novel datasets.
Several recent reviews focus on the usefulness of mathematical
modeling for understanding complex biological systems
\cite{Servedio.pb14, Mobius.c15, Torres.fg15, Karr.com15}.
Specifically, \citet{Mobius.c15} illustrated how mathematical modeling
helps to interpret experimental data using example of molecular
motors. While \citet{Mobius.c15} do describe some pros and cons of
mathematical modeling, there was limited discussion on what makes a
good mathematical modeling paper and when mathematical modeling can
lead to robust conclusions.

The authors did not cite and did not discuss an important
philosophical paper on the use of mathematical modeling in natural
sciences, a paper that every mathematical modeler should read, think
about, and discuss with his/her peers \cite{Oreskes.s94}. In contrast
with prevailing view in the mathematical modeling community (e.g.,
\cite{Mobius.c15}), \citet{Oreskes.s94} argued that for open natural
systems (and most if not all systems in biology are open) mathematical
models cannot make robust predictions. Moreover, based on
philosophical definitions the authors argued that verification of
mathematical models is impossible and validation of models, as
commonly used in the literature, is nearly impossible
\cite{Oreskes.s94}. They further argued that the use of mathematical
models should be rather heuristic, to understand limits of specific
mechanisms when models are constrained by experimental data.  If most
model predictions are not robust \cite{DeBoer.pcb12}, how should one
proceed with best use of mathematical modeling?

One approach, which was not strongly emphasized by \citet{Mobius.c15}
is the ``method of multiple working hypothesis'' by
\citet{Chamberlin.s1890} augmented by the ``strong inference'' of
\citet{Platt.s64}. In essence, instead of developing a single
mathematical model for the phenomenon in question as it appears to be
the norm in the field, one must consider several alternative models
and use experimental data as a judge in discriminating between
alternative hypotheses \cite{Platt.s64, Hilborn.b97}. In cases when
data are not available for model discrimination, analysis of
alternative models can inform which experiments may need to be
performed to discriminate between alternative models. Furthermore,
analysis of multiple alternative models could also highlight
limitations on predictions of any individual model, thus, testing
robustness of the model predictions which for any given model a priori
are expected to not be robust \cite{Oreskes.s94, Kirk.s15}.
Discrimination between alternative models using experimental data
allows to reject some of these models and rejection, rather than
confirmation of models is the strongest application of the scientific
method \cite{Platt.s64, Popper.b02}. Indeed, as \citet{Oreskes.s94}
highlighted ``models which challenge existing theories are more
valuable than models which confirm them.'' Confirmation of simplest
possible models is reasonable but showing that simple models do not
work is clearly more illuminating highlighting our gaps in
understanding the system in question \cite{Noecker.v15}.

As mathematical modeling is taking an integral part of biology it is
time that education and training of mathematical modelers is moving
beyond development and analysis of single models. Rather, development
of multiple mathematical models which are tested against experimental
data and finding which mathematical models are not able explain
experimental data needs to become a norm for the mathematical modeling
publications in the 21$^{st}$ century.

\bibliography{references}


\end{document}